\documentclass[sigconf]{acmart}

\AtBeginDocument{%
  \providecommand\BibTeX{{%
    \normalfont B\kern-0.5em{\scshape i\kern-0.25em b}\kern-0.8em\TeX}}}

\setcopyright{acmcopyright}
\copyrightyear{2018}
\acmYear{2018}
\acmDOI{10.1145/1122445.1122456}

\usepackage[utf8]{inputenc}
\usepackage{graphicx}
\usepackage{tabu}
\usepackage{array}
\usepackage{caption}
\usepackage[toc,page]{appendix}
\usepackage{hyperref}
\usepackage{float}
\usepackage{amsmath}
\usepackage[linesnumbered,ruled,vlined]{algorithm2e}
\usepackage{graphicx} 
\graphicspath{ {images/} }
\usepackage{url, color, soul}
\usepackage{latexsym} 
\usepackage{multirow}
\usepackage{verbatim}
\usepackage{xcolor}
\usepackage{xargs}
\usepackage{listings}
\usepackage{enumitem}

\begin{document}

\title{Smart Crawling: A New Approach toward Focus Crawling from Twitter}

\author{Ahmad Khazaie}
\affiliation{%
  \institution{LRI, CentraleSup\'elec}
  \streetaddress{Paris-Saclay University}
  \city{Gif-sur-Yvette 91190}
  \country{France}
}

\author{Nac\'era Bennacer Seghouani}
\affiliation{%
  \institution{LRI, CentraleSup\'elec}
  \streetaddress{Paris-Saclay University}
  \city{Gif-sur-Yvette 91190}
  \country{France}
  }

\author{Francesca Bugiotti}
\affiliation{%
 \institution{LRI, CentraleSup\'elec}
  \streetaddress{Paris-Saclay University}
  \city{Gif-sur-Yvette 91190}
  \country{France}
  }

\renewcommand{\shortauthors}{Khazaie, Bennacer, Bugiotti}

\begin{abstract}
  Twitter is a social network that offers a rich and interesting source of information challenging to retrieve and analyze. Twitter data can be accessed using a REST API. The available operations allow to retrieve tweets on the basis of a set of keywords but with limitations such as the number of calls per minute and the size of results. Besides, there is no control on retrieved results and finding tweets which are relevant to a specific topic is a big issue. Given these limitations, it is important that the query  keywords cover unambiguously the topic of interest  in order to both reach the relevant answers and decrease the number of API calls.\newline
In this paper we introduce a new crawling algorithm called "Smart Twitter Crawling"  (STiC) that retrieves a set of tweets related to a target topic. In this algorithm we take an initial keyword query and we enrich it using a set of additional keywords that come from different data sources.  STiC algorithm relies on a DFS search in Twitter graph where each reached tweet is considered if it is relevant with the query keywords using a scoring, updated throughout the whole crawling process. This scoring takes into account the tweet text, hashtags and the users who has posted the tweet, replied to the tweet, been mentioned in the tweet or retweeted the tweet.
Given this score STiC is able to select relevant tweets in each iteration and continue by adding the relate valuable tweets. Several experiments have been achieved for different kind of queries, the results showed  that the precision increases compared to a simple BFS search. 

\end{abstract}



\maketitle

\section{Introduction}
During the recent years, micro-blogging become a source of current and topical news. Twitter\footnote{www.twitter.com} is one of the most popular micro-blogging service which brings together millions
of users and allows to publish and exchange short messages, known under the name of \emph{tweets}. Twitter was a pioneer in providing APIs to access public data since 2006\footnote{https://developer.twitter.com/en.html} and enabling applications to retrieve tweets using a set of keywords. However, there is no control on the retrieved tweets which are not always relevant.  Reaching relevant answers requires  multiple calls and filtering the retrieved results. 

There are several research works that have as objective searching relevant tweets according to a given query~\cite{gabielkov2014sampling,gouriten2014scalable,li2013towards,safran2012improving}. Researchers tried to improve the ways  based on features  such as hashtags, retweets and mentions to retrieve the most relevant tweets from Twitter.  For example, one basic and simple method is if a query just contains one word, they find the frequency of this word in different features of Twitter\cite{nazi2015walk}. Researchers also tried to use external information extracted from sources such as Wikipedia to enrich a query by finding the most similar words to  find the most related results to the given query \cite{guisado2016enrich}.

{In this paper, we exploited different external resources 
such as  \textit{Wordnet}, \textit{Wordnik}, \textit{DBPedia}, and \textit{New York Times(NYT)} to have the most complete set of  keywords
 similar to the original query keywords. We also defined a smart crawling algorithm based on the tweet's features to reach the most relevant ones as early as possible and going beyond Twitter limitations.   More precisely, we define a new crawling algorithm called \textbf{Smart Twitter Crawling}  (STiC) by considering two different aspects: (i) Enriching the original query  using external sources,  (ii)  crawling Twitter graph using a DFS search based on new scoring to reach the best nodes (tweets) related to the targeted topic.
To measure the relevance of a tweet, the algorithm assigns a score to  a tweet 
by taking into account  its content, its hashtags and the 
user who has relation with it, such as posting, replying, being mention or retweeting. Given this score we are able to select highly related tweets in each iteration and continue by adding the relate valuable tweets.


Different experiments have been achieved on tweets collected for different kind of query keywords to evaluate the precision of STiC algorithm. 
Thanks to our approach, compared to a simple BFS search, we increased the precision of related retrieved tweets up to 86\% for some queries. Also in case of number of retrieved results, we got significant improvement in compare with simple BFS model.

In this paper, we first present related work  in Section~\ref{sec:relatedwork}.  Then, we present in detail our  smart Twitter crawling approach including query enrichment and Twitter graph  crawling in Section~\ref{sec:approach}.
In section~\ref{sec:experiments}, we present and discuss our results using different queries. Finally,in Section~\ref{sec:conclusion} we  give our conclusions and perspectives. 

\section{Related Work}
\label{sec:relatedwork}

Even before the emerging of social media, crawling the Web pages has been a common practice~\cite{li2013towards}. Finding the Web pages related to one topic was one of the interesting approaches to study~\cite{safran2012improving}. The common applied methodology, used neural network and vector space models to compute the priority models\cite{safran2012improving}. Deligenti\cite{diligenti2000focused} in 2000, introduced a model for focused crawling based on context graph by assigning appropriate credits to documents. 
Also Safran and et al.,\cite{safran2012improving} at 2012 proposed a new approach to improve relevance prediction in focused Web crawlers. They chose Na\"{\i}ve Bayesian as the base prediction model and they used four relevant attributes to create their prediction model: URL, anchor text, surrounding texts, and parent pages. They extended the list of keywords related to a topic by using WordNet and extracted relevant initial seed URLs automatically by choosing the top k-URLs retrieved from Google, Yahoo and MSN search engines.
Gouriten \cite{gouriten2014scalable}, in 2014 introduced an adaptive, scalable and generic system for focused crawling to identify and optimized the relevant subsystems. 
Their algorithm was defined for focused Web crawling, topic-centered Twitter user crawl, deep Web siphoning through a keyword search, gossip peer-to-peer search and real-world social network to answer a query. 
Xinyue Wang and et al, in 2015\cite{wang2015adaptive} studied about finding a solution for crawling Microblog feeds in real time. They proposed an adaptive crawling model which extracts the hashtags from Twitter iteratively to achieve a list of relevant tweets to a query.
Cha and et al, \cite{cha2010measuring} have worked on how to find most influential users in Twitter and his results could be useful when it be used to complete the idea for topic-focused crawling. 
In 2010, Tianyi and at al., \cite{wang2010unbiased} proposed a method to unbiased crawling the Tweets based on Metropolis-Hasting Random Walk(MHRW) using USDSG in the new method. 
Rui and et al., in 2013 \cite{li2013towards} proposed a data platform to automatically monitor “target” tweets from the Twitter stream for any specific topic. They designed Automatic Topic-focused Monitor (ATM), which first samples tweets from the Twitter stream and second selects a list of keywords to track based on the samples. 
GabielKov and et al.in 2014\cite{gabielkov2014sampling}, were working on sampling techniques for studying OSN. They have two scenarios for sampling and they want to find the best technique for each of them: first, they are looking for most popular users; the second one is that they have an aim to obtain unbiased sample of users. 
They showed that the classical sampling methods are highly biased by high degree nodes. \cite{gabielkov2014studying}
In \cite{kwak2010twitter}, they proved that BFS will have a large bias when the number of requests to the API is limited. In RW, choosing the next node for visiting, depends on the degree of the node. They used USDSG (Unbiased sampling for directed social graphs) algorithm, proposed in \cite{wang2010unbiased}, which is a modification of RW and discards a random jump to a node with a probability proportional to the degree of the node and replace arcs with undirected edges.\newline
Selecting keywords to retrieve relevant documents have been studied in lots of academic researches. As we mentioned earlier, Safran\cite{safran2012improving} at 2012 used WordNet to extend the extracted word set. 
Rui and co in 2013 \cite{li2013towards} proposed ATM Framework to select keywords in a constrained optimization approach, which finds near optimal keywords with guarantee (e.g., keywords are not too specific) and considers two types of costs. Also, ATM updates keywords in iterations which monitor the Twitter stream continuously.
In 2015, Xinyue Wang and et al.,\cite{wang2015adaptive} reviewed the retrieved tweets to identify new keywords for automatically live event tweet collection, these new keywords were mostly based on the hashtags which was embedded inside the tweet. 
Gusiado and et al. in 2016 \cite{guisado2016enrich} presented a query rewriting cloud service. Their aim is solving the problem of vocabulary mismatch and topic inexperience of users. So, 
they proposed a method which offers a generic solution by analyzing the websites using Wikipedia and identifying the entities called ENRICH. \\

\section{Smart Twitter Crawling Approach}
\label{sec:approach}
Monitoring the set of tweets related to a target topic is an unsolved problem~\cite{congosto2017t}. In this section we present the Smart Twitter Crawling (STiC) approach  we defined as a solution to this problem. The figure \ref{fig:architecture} describes the overall of our approach. 
STiC algorithm  enriches  initial  keywords using external sources to query Twitter graph. It  builds an initial sub-graph providing related seeds. The crawling is then based on a DFS search and exploits each considered tweet's features to assign a score and to select the most relevant tweets to be crawled. The results of the crawl will be a sub-graph made by different crawled nodes and the edges between them. This sub-graph will be stores in the a graph database, which is Neo4j in our work.   Before going any further into details, we first present the  input and output data representation of STiC algorithm. 
 
\begin{figure}
  \includegraphics[width=\linewidth, height=70mm]{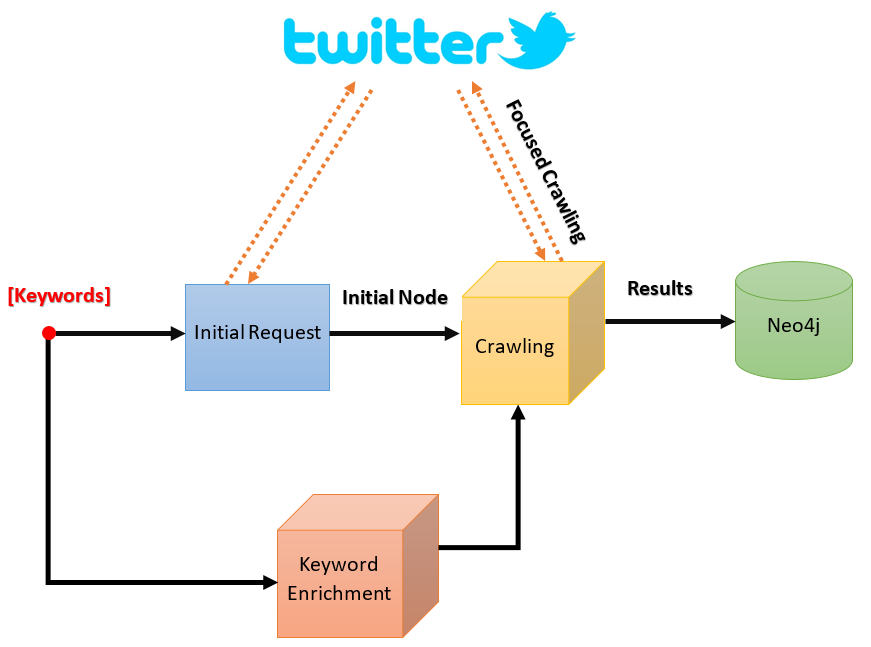}
  \caption{Architecture of our approach}
  \label{fig:architecture}
\end{figure}

\subsection{Input and Output  of STiC Algorithm}
\label{sec:representation}
Twitter data can be represented as a graph 
 $\mathcal{T} = <\mathcal{V} , \mathcal{U}> 
$
where $\mathcal{V}$ is the set of nodes and $\mathcal{U}$ is the set of directed edges between nodes. Different types of nodes are defined in $\mathcal{V}$:
\begin{itemize}
\item $t$ is a \emph{tweet},  accompanied by attribute values, which include the text of the tweet and its identifier.
\item $h$ is a \emph{hashtag} extracted from the tweet. 
\item $u$ is a \emph{user}  accompanied by its identifier value. 
\end{itemize}

 Different types of relations are defined in $\mathcal{U}$:
\begin{itemize}
\item \textbf{$<t,h>$} edge called $Has\_Hashtag$  which relates a tweet $t$  to a hashtag $h$ it contains. 

\item \textbf{$<t,t^\prime>$} edges called:
  \begin{itemize}
  \item $Quotes$ which relates a tweet $t$  to a tweet $t^\prime$. In this case, the text of node $t$ contains the text of $t^\prime$ in addition to its own text.
  \item $Replies\_To$  which relates a reply tweet $t$ to the origin tweet $t^\prime$. 
  \item $ReTweets$ which relates a retweet $t$ to the origin tweet $t^\prime$. In this case, the text of $t$ is exactly the same as text of tweet $t^\prime$.
  \end{itemize}

\item \textbf{$<t,u>$} called $Mentions$ which relates a tweet $t$  to the user  $u$  mentioned in it.

\item $\textbf{$<u,t>$}$ edges called:
	\begin{itemize}
	\item $Favorites$ which relates a user $u$ to a tweet $t$ which means $u$ likes $t$. 
    \item $Posts$ which relates a user  $u$ to a tweet $t$ which means $u$ posted $t$.
	\end{itemize}
\end{itemize}

\begin{itemize}
\item $\textbf{$<u,u\prime>$}$ edge called $Follows$ which relates a user $u$  to a user $u\prime$ he follows.
\end{itemize}

The input of STiC algorithm is the list of keywords after enrichment and an initial sub-graph of Twitter in which nodes has no any additional information than what is available from Twitter API. 
The output of the algorithm is a sub-graph, in which each node is accompanied with a score value. 

STiC algorithm defines 3 main procedures: 
\begin{itemize}
\item \textbf{Query enrichment procedure:} First step of algorithm to enrich list of keywords for the given query.
\item \textbf{Select new node procedure:} The procedure for selecting next node for being visited in crawling process.
\item \textbf{Smart crawling procedure:} The main process of algorithm to use the enriched list of keywords and node selection process for visiting new nodes and crawl Twitter.
\end{itemize}


\subsection{Query enrichment} 
\label{sec:queryEnrichment}
The REST Twitter APIs offer the possibility to retrieve tweets using a set of keywords. When a user tries to retrieve tweets he is not always conscious of the best set of keywords to use in order to obtain the correct subset of tweets. In our research we use the external sources to enrich the set of keywords that the user specifies in the target query. Alg.\ref{alg:CrawlAlgorithmQE} expresses the process of enriching a query. In this procedure, we collect all related words from different data sources, such as NYT~\cite{zhao2011comparing}, Wikipedia~\cite{guisado2016enrich} and WordNet~\cite{safran2012improving} APIs. We identified a list of APIs that provide as source of information  news, articles or synonyms and we identified the following ones: \textit{New York Times(NYT)}, \textit{Wordnet}, \textit{Wordnik}, \textit{DBPedia}, \textit{Thesaurus}, \textit{Datamuse}, \textit{WordsAPI}, \textit{Pydictionary}.

We also give a weight to each keyword to specify the relevance of the keyword to the original query keywords. For assigning this weight, we consider the subset of keyword given by each external source as a separated set and we calculate the IDF of each keyword. For this aim, each external source is considered as a document and then we calculated term frequency as the number of occurrence of the word in all documents. Then we assign the weight of each word as term frequency in all documents multiply by its IDF score.
For instance for \textit{obama} original query keyword we retrieve the following terms and their frequency:
({\textit{barack obama}: 4, \textit{barack hussein obama}: 3, \textit{barack}: 3, \textit{obama}: 3, \textit{community organizer in chief}: 2, \textit{barak obama}: 2, etc.}) \\

Then  the weight (total term frequency*IDF score) for each term is computed: 
\textit{barack obama}: 0.96, \textit{barack hussein obama}: 0.89, \textit{community organizer in chief}: 0.78, \textit{barak obama}: 0.78, etc.})\\

Finally we select the top score keywords:
(\textit{barack obama}, \textit{barack hussein obama}, \textit{community organizer in chief}, \textit{barak obama}, \textit{barrack obama}, \textit{president obama}, \textit{barackobama}, \textit{brack obama}, \textit{obama barack}, etc.) 

We finally merge the all keywords extracted from all APIs with their calculated weights and we sort them based on their weights and  on a threshold $\alpha$. The key factor for selecting $\alpha$ is that the keywords with the score above the threshold should not be irrelevant to the query. This threshold may vary and depends on the type of the query and results of the query enrichment. 
The algorithm \ref{alg:CrawlAlgorithmQE} describes the function \textit{mostRelatedKeywords} which returns the 

\begin{algorithm}[t]

\DontPrintSemicolon
\SetKwInput{KwInput}{Input}                
\SetKwInput{KwOutput}{Output}              
\SetKwProg{Fn}{Function}{:}{}
\caption{STiC Algorithm - Query Enrichment}
\label{alg:CrawlAlgorithmQE}

\KwInput {$external\_sources\_list$ \tcp*{\textit{New York Times(NYT)}, \textit{Wordnet}, \textit{Wordnik},\textit{DBPedia}, \textit{Thesaurus}, \textit{Datamuse}, \textit{WordsAPI}, \textit{Pydictionary}}}
\KwInput {$\alpha$ \tcp*{\textit{keyword relevance threshold}}} 
\KwInput {$query$}
\KwOutput{$keywords\_list$}
\BlankLine

\For {$source$ in $external\_sources\_list$}
{
      {$keywords\_list.add(related\_words(source, query))$}
}
{$calculateTFIDFScore(keywords\_list)$}\\
{\textbf{return} {$mostRelatedKeywords(keywords\_list,\alpha)$}}

\end{algorithm}

\subsection{Smart Crawling} 
\label{sec:smartCrawling}

In the Alg.\ref{alg:CrawlAlgorithmSC}, the list of keywords from the Alg.\ref{alg:CrawlAlgorithmQE}, the initial node and number of iterations, will be given to the procedure and in first iteration, a sub-graph of nodes from neighbors of initial node will be created and scores of nodes will be updated. 
The initialization of the graph is crucial for the success of the following iterations of the algorithm: a bad initialization can lead to a graph where there is any node is related to query. In the beginning, we chose manually a single node we knew was relevant (e.g. we can take the official hashtag if it is in the set of keywords specified by the user). While quite effective, this selection cannot be transparently done since it needs manual selection for each different query and there is chance of leading the crawling to a specific part of data.
We initialize the graph automatically with the result of a simple Twitter API search using the enriched keywords. This preliminarily results allow for the first round of the crawl.
Then, as number of iterations, the crawler will visit the selected node from Alg.\ref{alg:CrawlAlgorithmNS}, which is explained in \ref{sec:nodeSelection} and it will add its neighbors to the list of candidate nodes, which will be given to the Alg.\ref{alg:CrawlAlgorithmNS} in the next iteration. Then, before going for the next iteration, we need to update the scores of nodes which is explained in \ref{sec:scoreCalculation}. In the end, a sub-graph of Twitter will be returned which has been created by crawling nodes based on the defined score for each one.

\begin{algorithm}[t]
\DontPrintSemicolon
\SetKwInput{KwInput}{Input}                
\SetKwInput{KwOutput}{Output}              
\SetKwProg{Fn}{Function}{:}{}
\caption{STiC Algorithm - Smart Crawling}
\label{alg:CrawlAlgorithmSC}
 
\KwInput {$keywords\_list$}
\KwInput {$iterations$}
\KwInput {$initial\_relevant\_node$}
\KwOutput{$visited\_nodes$ }
\BlankLine

{ $i \gets \textit{iterations}$ }\\
{ ${ n_0 \gets \textit{initial\_relevant\_node} }$ }\\
{ $current\_node \gets n_0$ }\\
{ $update\_queue\_nodes(current\_node\_neighbors)$ }\\
{ $update\_nodes\_scores()$ }\\
{ $add\_to\_visited\_list(current\_node)$ }\\
{ $\textit{i} \gets \textit{i - 1}$ }\\

  \While {i $>$ 0}
  {
      { $current\_node \gets \textit{selectNewNode(queue\_nodes)}$ }\\
      \While{$is\_visited(current\_node)$}
      {
          { $current\_node \gets \textit{selectNewNode(queue\_nodes)}$ }\\
      }
      { $update\_queue\_nodes(current\_node\_neighbors)$ }\\
      { $update\_nodes\_scores()$ }\\
      { $add\_to\_visited\_list(current\_node)$ }\\
      { $\textit{i} \gets \textit{i - 1}$ }\\
  }
  \textbf{return} $\textit{visited\_list}$
\end{algorithm}

\subsection{Node Selection}
\label{sec:nodeSelection}
Our crawling method selects the next node from which to continue the crawling during each iteration. On the one hand we want to explore the most promising nodes, i.e. the ones with the highest estimate scores, and not waste queries on irrelevant nodes. On the other hand, we would also like to avoid remaining in a portion of the Twitter graph and miss other relevant nodes. The first objective can be understood as efficiency whereas the second as completeness. A common solution to this trade off is the introduction of random based choice. In the equations below, the $p$ (a real number between 0 and 1) parametrizes the probability distribution. The closer $p$ is to 1, the higher the probability to chose a high score is. On the contrary, if $p$ is close to 0, the low scored nodes will have a higher probability of being chosen. This probability distribution is inspired by a multinomial distribution and a soft-max function. For each node $i$, the probability to be selected $P_i$ is given by the non-normalized multinomial function $f_i$, and depending of the parameter $p$:
	
	\[
	P_i = \dfrac{ \exp \left( f_i \right)}{\underset{i}\sum  \exp \left( f_i \right)}
	\]

	\[
	\text{where: }f_i =  \dfrac{x_i }{x_{\min}} \cdot p + \dfrac{x_{\max}}{x_i} \cdot (1-p)
	\]

Using this formula, we are able to jump from one node to another one if the score of the node in not large enough to be selected for crawling. In this case, we use the minimum and maximum scores of the crawled nodes and choose $p$ arbitrary, to define the function $f_i$ for each node. The probability of $P_i$ shows the chance of a node to be selected based on the fraction of its $f_i$ to sum of $f_i$ for other crawled nodes.
In next section we are going to describe the process of calculating the score for different types of nodes. The algorithm \ref{alg:CrawlAlgorithmNS}

\subsubsection{Score Calculation}
\label{sec:scoreCalculation}


At the beginning of each iteration, a node, \textit{$n_0$}, is selected and all internal information about this node is retrieved from Twitter which includes the \textit{id\_str} of its neighbors, who are then added to the queue for the next iteration. Then, the scores of all nodes are updated. For \textit{$n_0$} only \textsc{text\_score} is available at the beginning and as there is no other node which has been visited before that, its \textsc{estimate\_score} is equal to 0. The final score is calculated according to the various score attributes to find the relevance of a node according to the initial query. 

The score related to a text of a tweet is  defined as follows:
\begin{itemize}
\item \textsc{text\_score}$(t)$: is  defined for a tweet  node $t$  and is  represent the frequency of query  keywords in the text body of the tweet.
\subsubsection{Tweets content analysis}
\label{sec:contentAnalysis}
Contrary to the User and Hashtag  nodes (the Hashtag nodes are merely a
word or a name), a tweet is characterized by a textual content that allows us to use Natural Language Processing tools to judge their relevance to the target topic. We begin this step with a list of keywords. The analysis of the tweet consists in a lexical and semantic comparison between the keywords and the text body. This analysis begins with the lemmatization of both texts. 
This is a classic NLP tool that transforms the words into their root form. This allows us to ignores plurality for nouns and tenses for verbs. Punctuation marks and linking words (e.g. the, and, a, of . . . ) are removed because they usually do not convey useful semantic knowledge. Both texts are then compared both lexically and semantically. The lexical comparison is done by counting the number of words the texts have in common. We note that this count is not normalized, but the limit of 280 characters of a tweet prevents the possibility of a longer text that contains a lot of keywords. The semantic comparison is done using the Word Net database. In this database, words possess various relationships with each other. In particular, we utilize the hyponym relationship:
the link between two words is be measured as the depth of the closest common ancestor in the hyponymy graph. A keyword is considered to match a word with a semantical relation if the similarity value given by Word Net is higher than a threshold set beforehand. At last, the score from the text of a tweet is the sum of weights of keywords matched (either by semantic relation or lexical).\\

\item \textsc{estimate\_score}: is estimating the relevance of the node, $n\in$ $\mathcal{V}$ based on the score of a direct predecessor node, $n^\prime\in$ $\mathcal{V}$, which is a visited node that has a relation with $n$ and the edge $e\in$ $\mathcal{U}$ connects $n$ and $n^\prime$ together.
  \begin{itemize}
  \item {for a Tweet}: \textsc{tweet\_estimate\_coef}$ = [0.4, 0.6, 1.0, 1.0, 1.0, 0.5,0.5]$\\ These coefficients concern, in order, the user who posts the tweet, mentioned users in this tweet, original of this tweet if it replies to another one, original of this tweet if it quotes another one, original of this Tweet if it is a retweet of another one, and retweets of this tweet.
  \item {for a User}: \textsc{user\_estimate\_coef}$ = [1.0, 0.6, 0.5, 0.3]$\\
  These coefficients concern, in order, tweets posted by this user, his favorite tweets, his friends, and his followers.	
  \end{itemize}
\item \textsc{feedback\_score}: is estimating the relevance of the node, $n\in$ $\mathcal{V}$ based on the score of direct successor nodes $n^\prime\in$ $\mathcal{V}$, which is the one who has been visited after $n$ and there is edge $e\in$ $\mathcal{U}$ between $n$ and $n^\prime$ to show the relation between nodes. 
\item \textsc{score}: this final score is computed after the crawling and feedback steps of the algorithm and it is calculated based on the three previous scores.
\begin{itemize}
\item {for a Tweet}:   $Score$ = $text\_score$ + $feedback\_score$
\item {for a User}:    $Score$ = $estimate\_score$ + $feedback\_score$
\item {for a Hashtag}:    $Score$ = $estimate\_score$ + $feedback\_score$ + $Occurance\_Count$
\end{itemize}

\end{itemize}
	
To obtain a node's {\sc estimate\_score}, we multiply it's predecessor's {\sc score} by the corresponding coefficient.

Thus, a tweet node has 4 score-related attributes whereas other node types  have 3. These attributes exist regardless of the node's state. We assume at the start that we begin with some seed tweets, considered highly relevant. The precise way in which we obtain those tweets is detailed in Section 3.4. We evaluate the \textsc{text\_score} of these seeds using the strategy described in Section 3.2. We set their \textsc{estimate\_score} equal to their \textsc{text\_score} to allow our algorithm to run. At each iteration during the crawling, we begin by selecting a new node using the method described in Section 3.5. We then query Twitter to complete its information. We update the \textsc{score} of the nodes as follows: if it is a tweet, we compute its \textsc{text\_score} and We add the difference between calculated \textsc{text\_score} and its \textsc{estimate\_score} to its parent's \textsc{feedback\_score}. We then proceed to add this node's uncrawled neighbors to the graph. We set their \textsc{estimate\_score} as a fraction of the current node's score, based on the relationship they share. If it is a User node, the score will be equal to sum of  its \textsc{estimate\_score} and \textsc{feedback\_score}. If it is a Hashtag node, in addition to sum of its \textsc{estimate\_score} and \textsc{feedback\_score}, we count how often they appear and add it to their score.

Alg.\ref{alg:CrawlAlgorithmNS} defines the process of selecting a new node. In this procedure, the input is a list of candidate nodes and for each node in this list, the function $f$ will be calculated and then selecting probability for it, $P$ will be calculated. In the end, the node with the highest probability will be returned. 

\begin{algorithm}[t]
\DontPrintSemicolon
\SetKwInput{KwInput}{Input}                
\SetKwInput{KwOutput}{Output}              
\SetKwProg{Fn}{Function}{:}{}
\caption{STiC Algorithm - Node Selection}
\label{alg:CrawlAlgorithmNS}

\KwInput {$queue\_nodes$}
\KwInput {$ p \gets 0.7$ \tcp*{probability of selecting high score node}}
\KwOutput {$selected\_node$}
\BlankLine

{ $max\_score \gets \textit{maximum score of queue nodes}$ }\\
{ $min\_score \gets \textit{minimum score of queue nodes}$ }\\
  \For {$node$ in $queue\_nodes$}
  {
      { ${ f[i] = calculate\_F(node.score, min\_score, max\_score, p) }$ }\\
      { $P[i] = exp(f[i])/sum(exp(f[i]))$ }\\
  }
  { \textbf{return} {$node\_with\_max\_P[i]$} }\\

\end{algorithm}

\section{Experiments and Evaluation}
\label{sec:experiments}

STiC algorithm is implemented by Python, we used tweepy\footnote{http://www.tweepy.org/} v3.5 to access the Twitter API, and neo4j-driver\footnote{https://neo4j.com/developer/python/} v1.0.2 and neo4jrestclient\footnote{https://pypi.org/project/neo4jrestclient/} v2.1.1 to communicate with Neo4j.
For enriching the list of keywords we used different APIs and all of them are implemented in python\footnote{https://www.python.org/download/releases/3.4.0/} 3.4. In some cases we needed to create a new library while for others used predefined libraries.

We aimed to increase the precision of  retrieved tweets. In order to evaluate our approach to see how much STiC is successful, we run the experiments with maximum 100 iteration for crawling and maximum timeout 720 second. The relevance threshold for keywords, $\alpha$, is chosen as equal as 0.5 and the threshold for selecting high score node, $p$, is 0.7. These numbers are arbitrary and selected after observing a few iterations of crawling.
We run the model on each query separately and stored the results to be able to compare them by statistics and manual check.

We selected four original queries from four different categories including proper nouns: \textit{obama}, general words: \textit{bank}, concepts: \textit{energy} and recent trends: \textit{birlinggap}.  The reason for choosing these keywords is covering different categories of queries and being able to evaluate the system with different inputs and decrease the bias toward the specific part of tweets or users. \textit{obama} is the previous president of United States and he has huge number of followers, hashtags, mentions and tweets and it is very good option to start the crawling. 'bank' and 'energy' are very general and they have good number of relations and hashtags in Twitter. Also there is a lot of number of users which has significant number of related tweets and we can have a good chance to crawl an enough large subset of the crawling space. \textit{birlinggap} was one of the recent trends at the moment of doing experiments and it gives us the chance to do manual check on results easily. 

\begin{figure}[t]
 	\includegraphics[width=0.85\linewidth, height=75mm]{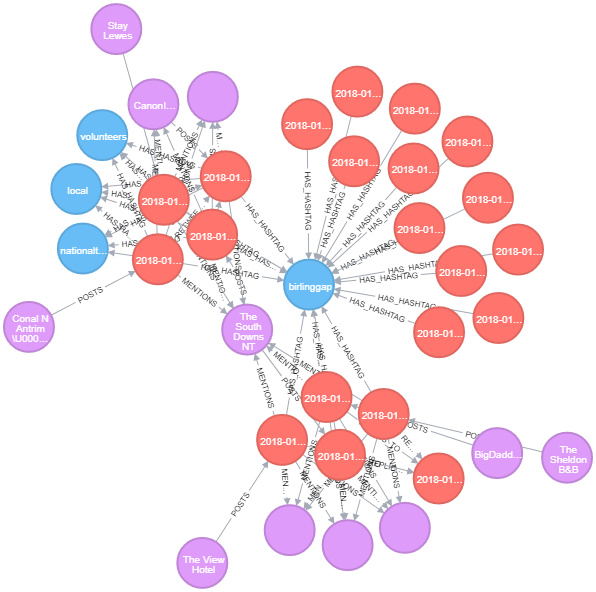}
    \caption{Retrieved nodes for query \textit{birlinggap} using STiC}
    \label{fig:birlinggap-new}
\end{figure}

Fig.\ref{fig:birlinggap-new} shows the retrieved nodes using STiC after storing in the database. Red nodes represent tweets in the database while blue nodes show the hashtags which found related to the query and purple nodes indicate users which has been crawled during the process. The edges' labels define the type of relation between nodes.

\begin{figure}[t]
    \includegraphics[width=\linewidth]{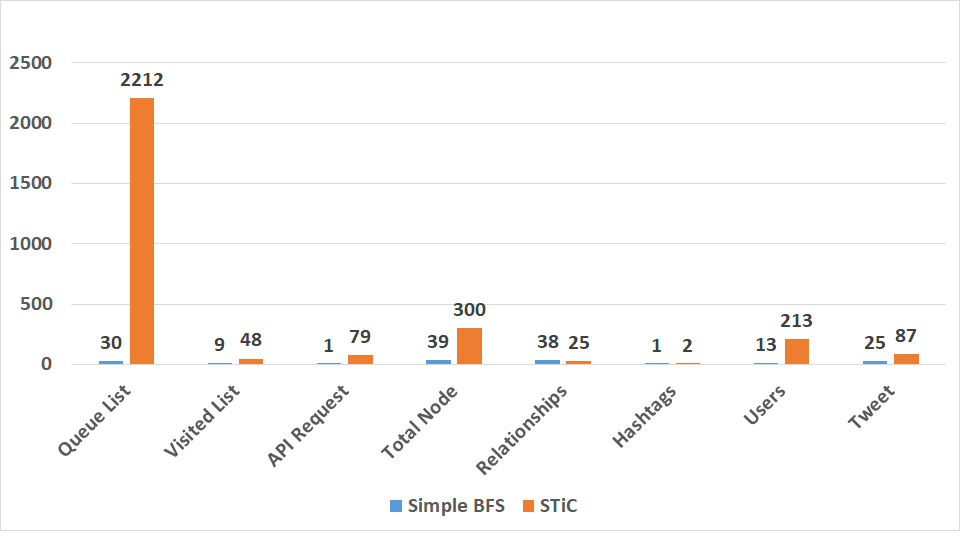}
    \caption{Number of different nodes for query 'obama'}
    \label{fig:obama-nodes}
\end{figure}

\begin{figure}[t]
 	\includegraphics[width=\linewidth]{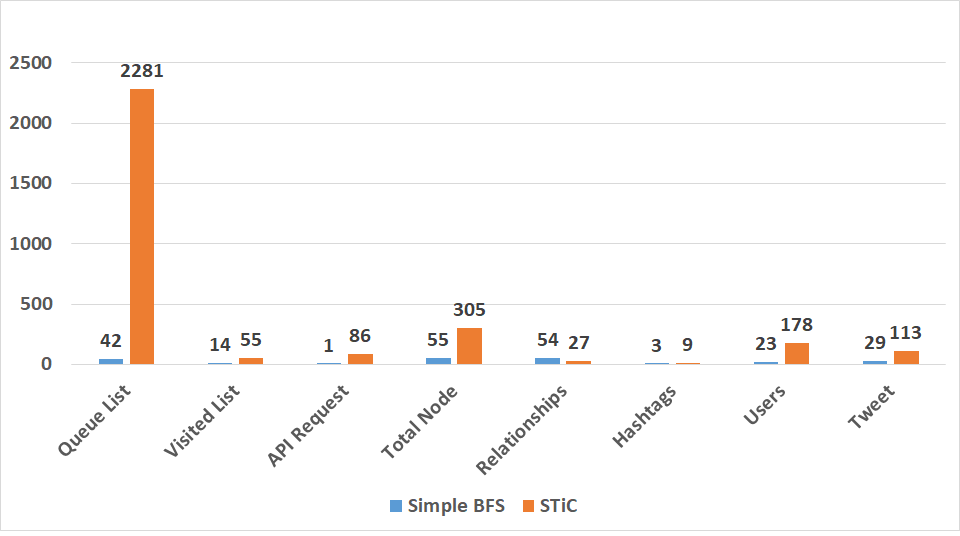}
    \caption{Number of different nodes for query 'energy'}
    \label{fig:energy-nodes}
\end{figure}

Figure \ref{fig:obama-nodes} and Figure \ref{fig:energy-nodes} show the number of different types of nodes for queries \textit{obama} and \textit{energy} using STiC, respectively and compare them with results of simple BFS API call for the given query. STic used more API calls and found more related tweets. It decreased the variety of relations between nodes and brought more Hastag and User node in compare to simple BFS method.


\begin{figure}[t]
    \includegraphics[width=\linewidth]{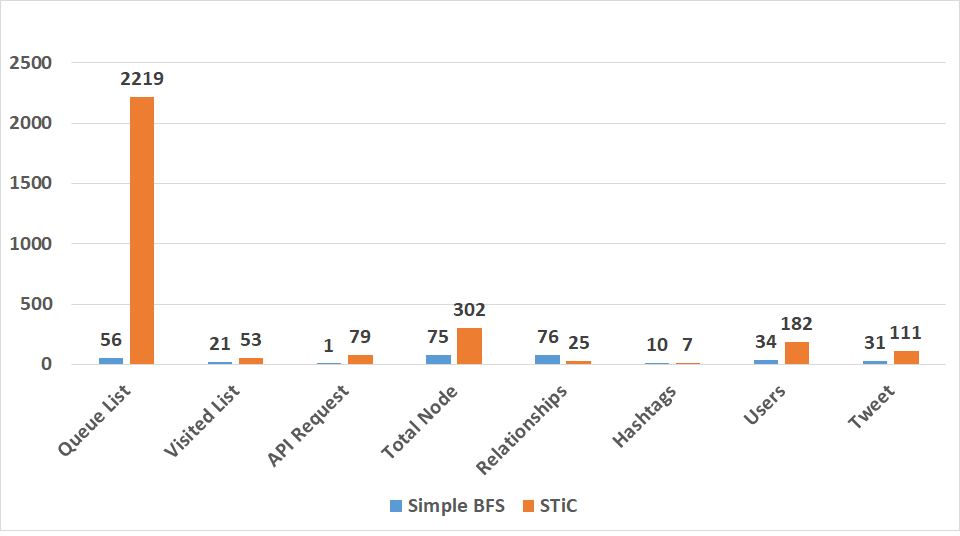}
    \caption{Number of different nodes for query 'bank'}
    \label{fig:bank-nodes}
 \end{figure}

Figure \ref{fig:bank-nodes} gives the comparison for different nodes retrieved by STiC and simple BFS for query \textit{bank}. STiC gives more tweet nodes and more User nodes but the number of Hashtag nodes are less than simple method. The reason is that STiC build the connected graph by crawling more users and tweets rather than jumping from node to another one without any relation.

\begin{figure}[t]
 	\includegraphics[width=\linewidth]{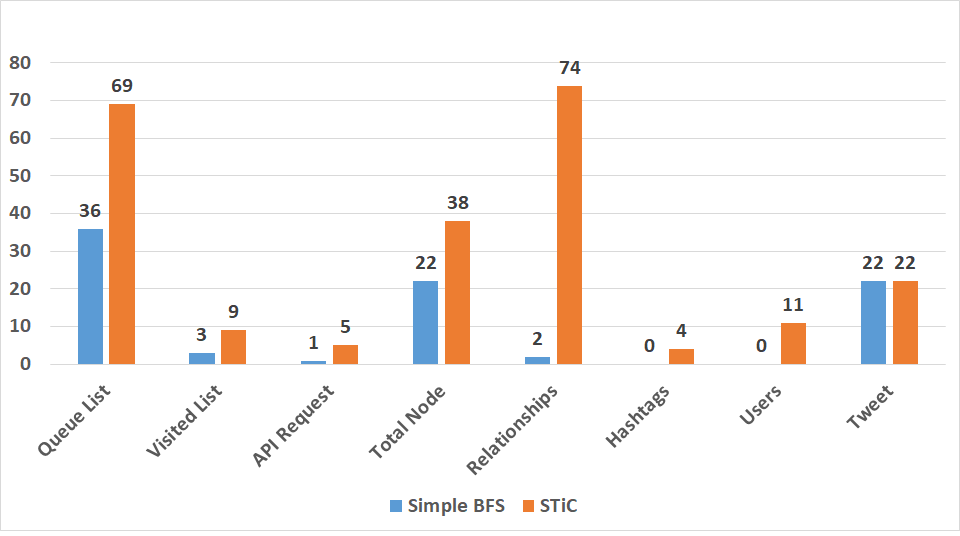}
    \caption{Number of different nodes for query 'birlinggap'}
    \label{fig:birlinggap-nodes}
\end{figure}

Figure \ref{fig:birlinggap-nodes}, shows that STiC found the same number of tweets for \textit{birlinggap} query while it is using more API calls and increase number of relationships between nodes. This increment is explained by comparing number of Hashtags and Users, since it found more nodes than simple BFS, these nodes caused more edges than before.

In this part for final evaluation, we decided to calculate precision for both STiC and simple BFS API call and see the improvement clearly.

in the tables \ref{table:Measures-obama}, \ref{table:Measures-birlinggap}, \ref{table:Measures-energy} and \ref{table:Measures-bank} you can find the precision percentage for simple BFS API call model and STiC model for each topic. STiC model shows significant improvement in finding related tweets, specially in \textit{birlinggap} query in which the simple BFS could not find any Hashtag or User nodes, which is very important for building relations and making connections between nodes.

\begin{center}
\captionof{table}{Precision result comparison for 'obama'}
\label{table:Measures-obama}
\begin{tabu}to 0.95\linewidth{ | X[l] |X[c] |X[c] |X[C]|}
 \hline
 \textsc{} & Retrieved Relevant Tweet & Retrieved Tweet & Precision\\
 \hline
 \textit{STiC} & 71 & 87 & 81.6\% 
\\
 \hline
 \textit{Simple BFS} & 16 & 25 & 64\% 
 \\
 \hline
\end{tabu}
\end{center}

\begin{center}
\captionof{table}{Precision result comparison for 'birlinggap'}
\label{table:Measures-birlinggap}
\begin{tabu}to 0.95\linewidth{ | X[l] |X[c] |X[c] |X[C]|}
 \hline
 \textsc{} & Retrieved Relevant Tweet & Retrieved Tweet & Precision\\
 \hline
 \textit{STiC} & 15 & 21 & 71.43\% 
\\
 \hline
 \textit{Simple BFS} & 8 & 22 & 36.36\% 
 \\
 \hline
\end{tabu}
\end{center}

\begin{center}
\captionof{table}{Precision result comparison for 'energy'}
\label{table:Measures-energy}
\begin{tabu}to 0.95\linewidth{ | X[l] |X[c] |X[c] |X[C]|}
 \hline
 \textsc{} & Retrieved Relevant Tweet & Retrieved Tweet & Precision\\
 \hline
 \textit{STiC} & 89 & 113 & 78.76\% 
\\
 \hline
 \textit{Simple BFS} & 16 & 29 & 55.17\% 
 \\
 \hline
\end{tabu}
\end{center}

\begin{center}
\captionof{table}{Precision result comparison for 'bank'}
\label{table:Measures-bank}
\begin{tabu}to 0.95\linewidth{ | X[l] |X[c] |X[c] |X[C]|}
 \hline
 \textsc{} & Retrieved Relevant Tweet & Retrieved Tweet & Precision\\
 \hline
 \textit{STiC} & 96 & 111 & 86.49\% 
\\
 \hline
 \textit{Simple BFS} & 24 & 31 & 77.42\% 
 \\
 \hline
\end{tabu}
\end{center}

Based on the achieved results, which have been shown in the tables and figures above, we observed the significant improvement in term of precision, number of retrieved tweets and different types of nodes after using STiC algorithm. This method shows high performance for crawling the Twitter and finding related tweets for a given query. In compare to simple BFS API call, STiC is able to retrieve more related tweets, while it is finding more Hashtags and Users and extending list of nodes during crawling Twitter. This method use more API calls since it can find stronger relation between visited nodes and uncrawled ones. For some queries such as \textit{birlinggap} which is a proper noun and there is small set of related words for them, simple BFS API call can not reach to a well connected graph and most of the nodes are not connected to each other while the STiC can build a graph with more edges between the nodes. For other queries, STiC can build a connected graph with more nodes and less diversity in number of relationships.
By comparing the results of STiC model for all queries, we observed that the number of queries having more related keywords how also a greater number of related nodes with respect to the queries having a smaller set of keywords.

\section{Conclusion and perspective}
\label{sec:conclusion}
In this paper, we aimed at developing a system for crawling relevant tweets to a given topic using a keyword query. We considered two aspects of the problem: the keyword-set enrichment and the crawling of relevant tweets using the Twitter APIs. First we focused on enriching queries and we used different external APIs (WordsAPI, Datamuse, Thesaurus, DBPedia, Wordnik, PyDictionary, Wordnet and New York Times API) and identified related keywords. We calculated TF-IDF score for these keywords and we removed the ones with lower score than threshold. We claimed that we can get more related tweets while we use more related keywords for a given topic. In the second step we defined a crawling algorithm and a scoring system in which each tweet is annotated by a score. Our crawling algorithm takes advantage of the text content of tweets using Natural Language Processing tools to deduce the relevance of each node. Overall, we obtain very satisfying results on well known topics, as a large number of retrieved tweets are related to the topic and number of retrieved tweets for running the model in a short period of time seems to be enough. Twitter is dynamic, as several thousands of tweets are posted each second, and the Twitter graph is in constant evolution. However the solution we developed seems to be resilient to these changes. 
This work opens the door for further interactions between various data
sources. We could also consider taking advantage of more than just the concepts from the APIs (e.g. the content of the articles). We would also have liked to test this process on a larger number of iterations, but we were limited by the manual aspect of our evaluation method.
For future work, we are going to improve the performance of finding related new tweets by using machine learning algorithms. So we will try to build a supervised learning system to classify new tweets by using the collected tweets as a train set. In this case we are able to use this system in many cases that have overlap with train set and sample queries. For being more precise and having significant improvement in pruning unrelated tweets, we can use the idea of popular users and most effective users to improve the the performance of scoring system and giving weight to the users as well. Another idea for smart crawling the tweets is using the provided URL in tweets and applying NLP methods on the text of the Web pages besides considering the meta data of the Web pages to create a more related list of keywords and hashtags for crawling the new tweets.

\section{Acknowledgment}
\label{sec:acknowledgement}
Our contributions consist in (i) the usage of several data sources in order to enrich a keyword query; (ii) the definition of a Smart Crawling algorithm as the main method to access the related tweets to an original query using the keyword enriched query and the REST Twitter APIs. We would like to thank V. Chetyrkine, C. Hamelain, X. Li for their work in the development of the tool which was the base model for STiC and Benoit Grotz and Silviu Maniu for many helpful discussions.


\bibliographystyle{ACM-Reference-Format}
\bibliography{main}

\end{document}